\documentclass[reprint,aps,prd,superscriptaddress,nobibnotes,nofootinbib,floatfix,showpacs,showkeys]{revtex4-2}
\usepackage{graphicx}

\usepackage{amsfonts}
\usepackage{mathrsfs}
\usepackage{amsmath}
\usepackage{amssymb}
\usepackage{framed}
\usepackage{multirow,tabularx}
\usepackage{bm}
\usepackage{orcidlink}
\usepackage[normalem]{ulem}
\usepackage{extarrows}
\usepackage{slashed}
\usepackage{isodateo}
\usepackage{booktabs}
\usepackage{xcolor}  % 可选，用于自适应列宽
\hypersetup{
	colorlinks=true,
	citecolor  =red,
	urlcolor=blue
}
\makeatletter
\let\SCGE@revtexabstract\abstract
\let\SCGE@revtexendabstract\endabstract
\renewcommand{\abstract}[1]{\SCGE@revtexabstract#1\SCGE@revtexendabstract}
\makeatother
\usepackage{mdframed}

\newmdenv[skipabove=0pt,%
skipbelow=5pt,%
leftmargin=0pt,%
rightmargin=0pt,%
innertopmargin=-5pt,%
innerbottommargin=7pt,%
innerleftmargin=2pt,%
innerrightmargin=2pt,%
splittopskip=0pt,%
splitbottomskip=0pt,%
linewidth=0pt,%
nobreak=true]%
{keyeqn2}

\newmdenv[backgroundcolor=gray!15,%
skipabove=0pt,%
skipbelow=5pt,%
leftmargin=0pt,%
rightmargin=0pt,%
innertopmargin=-5pt,%
innerbottommargin=7pt,%
innerleftmargin=2pt,%
innerrightmargin=2pt,%
splittopskip=0pt,%
splitbottomskip=0pt,%
linewidth=0pt,%
nobreak=true]%
{keyeqn}

\begin{document}

	\title{Exact Black Hole  Solutions in Bumblebee Gravity with Lightlike or Spacelike VEVs}

	\author{Jia-Zhou Liu\,\orcidlink{0009-0005-5430-4258}}
	\affiliation{Lanzhou Center for Theoretical Physics, Key Laboratory of Theoretical Physics of Gansu Province, Key Laboratory of Quantum Theory and Applications of MoE, Gansu Provincial Research Center for Basic Disciplines of Quantum Physics, Lanzhou University, Lanzhou 730000, China}
	\affiliation{Institute of Theoretical Physics $\&$ Research Center of Gravitation, School of Physical Science and Technology, Lanzhou University, Lanzhou 730000, China}

\author{Shan-Ping Wu\,\orcidlink{0000-0001-6633-1054}}
\affiliation{Lanzhou Center for Theoretical Physics, Key Laboratory of Theoretical Physics of Gansu Province, Key Laboratory of Quantum Theory and Applications of MoE, Gansu Provincial Research Center for Basic Disciplines of Quantum Physics, Lanzhou University, Lanzhou 730000, China}
\affiliation{Institute of Theoretical Physics $\&$ Research Center of Gravitation, School of Physical Science and Technology, Lanzhou University, Lanzhou 730000, China}

\author{Shao-Wen Wei\,\orcidlink{0000-0003-0731-0610}}
\affiliation{Lanzhou Center for Theoretical Physics, Key Laboratory of Theoretical Physics of Gansu Province, Key Laboratory of Quantum Theory and Applications of MoE, Gansu Provincial Research Center for Basic Disciplines of Quantum Physics, Lanzhou University, Lanzhou 730000, China}
\affiliation{Institute of Theoretical Physics $\&$ Research Center of Gravitation, School of Physical Science and Technology, Lanzhou University, Lanzhou 730000, China}

	\author{Yu-Xiao Liu\,\orcidlink{0000-0002-4117-4176}}
	\email{liuyx@lzu.edu.cn}
	\affiliation{Lanzhou Center for Theoretical Physics, Key Laboratory of Theoretical Physics of Gansu Province, Key Laboratory of Quantum Theory and Applications of MoE, Gansu Provincial Research Center for Basic Disciplines of Quantum Physics, Lanzhou University, Lanzhou 730000, China}
	\affiliation{Institute of Theoretical Physics $\&$ Research Center of Gravitation, School of Physical Science and Technology, Lanzhou University, Lanzhou 730000, China}

	%\date{}
	
	\abstract{Motivated by recent developments in Lorentz-violating theories of gravity, we obtain new black hole solutions within the framework of bumblebee gravity, where the bumblebee vector field possesses two independent nonzero components and acquires either a lightlike or spacelike vacuum expectation value. Within this framework, we derive new Schwarzschild-like and Schwarzschild–(A)dS-like black hole solutions. By further incorporating a nonminimally coupled electromagnetic field, we generalize these to new charged black hole solutions. These solutions extend previous results by including additional Lorentz-violating parameters. A key finding is that even for lightlike vacuum expectation values, the black hole solutions exhibit distinct corrections from Lorentz violation. Furthermore, we present a preliminary analysis of their thermodynamic properties. Similar to previous studies that reported a discrepancy between the black hole entropy and the Wald entropy in bumblebee gravity with spacelike vacuum expectation values, our solutions in the spacelike case exhibit the same behavior. In contrast, for the lightlike case considered here, the two entropies coincide.
		
	}
	

	\maketitle

	%==========================
\section{INTRODUCTION }
		Several candidate theories of quantum gravity have been proposed~\cite{Birrell:1982ix,Maldacena:1997re,Aharony:1999ti,Gubser:1998bc,Alfaro:1999wd,Alfaro:2001rb,Rovelli:1989za}. However, most quantum gravity effects are expected to occur only at the Planck scale ($\sim 10^{19}$ GeV), which lies far beyond the reach of current experimental techniques. Interestingly, a number of approaches predict that Lorentz symmetry may be violated in the IR regime of gravity, with possible signatures appearing at much lower, experimentally accessible energy scales.  
		
		The idea of spontaneous Lorentz symmetry breaking, first proposed in the context of string theory~\cite{Kostelecky:1988zi}, motivated the development of the Standard-Model Extension (SME) as a comprehensive framework for Lorentz-violating physics~\cite{Colladay:2001wk,Kostelecky:2000mm,Kostelecky:2001mb,Colladay:2009rb,Berger:2015yha,Carroll:1989vb,Andrianov:1994qv,Andrianov:1998wj,Lehnert:2004hq,BaetaScarpelli:2012kt,Brito:2013npa}. Within this framework, the so-called bumblebee model has been introduced as a simple but powerful toy model~\cite{Kostelecky:1989jw,Kostelecky:1988zi,Kostelecky:2003fs}, in which a self-interacting vector field $B_\mu$—the bumblebee vector—couples nonminimally to gravity. The vector field acquires a nonzero vacuum expectation value (VEV) through an appropriate potential, leading to spontaneous Lorentz-symmetry breaking in the gravitational sector.  In this scenario, the nonzero VEV of the bumblebee field selects a preferred direction in spacetime, thereby breaking local Lorentz symmetry spontaneously. Despite its conceptual simplicity, the bumblebee model captures essential features of Lorentz-violating gravity and has been widely used as an effective framework to explore possible phenomenological consequences of Lorentz symmetry breaking. In particular, the coupling between the bumblebee field and spacetime curvature can modify the gravitational dynamics and may lead to deviations from general relativity in strong gravitational fields.
		 However, more general configurations of the vector field may lead to richer structures of Lorentz symmetry breaking and potentially new gravitational effects. Such configurations may provide a broader setting to investigate how Lorentz violation influences the geometry and physical properties of gravitational systems.  The aim of this work is to derive exact black hole solutions in which the bumblebee vector field has two independent nonzero components, with the VEV being either lightlike or spacelike.
		
		Black holes provide an important arena for testing modified theories of gravity, especially in the strong-field regime. Recent observational advances have opened new possibilities to probe the properties of astrophysical black holes. For instance, the Event Horizon Telescope (EHT) has produced horizon-scale images of supermassive black holes~\cite{EventHorizonTelescope:2019dse,EventHorizonTelescope:2019ggy,EventHorizonTelescope:2019jqa,EventHorizonTelescope:2019pgp,EventHorizonTelescope:2019uob,EventHorizonTelescope:2019ggy6}, while gravitational-wave observations from compact binary mergers by LIGO–Virgo–KAGRA allow precise tests of gravitational dynamics in strong gravitational fields~\cite{Abbott:2016blz,Abbott:2016nmj,Abbott:2016nhf,Abbott:2018lct}. In addition, the orbital motion of S-stars around the supermassive black hole at the Galactic center provides further constraints on possible deviations from general relativity~\cite{Gillessen:2008qv,Gravity:2018ofz,Gravity:2020fml}. Therefore, studying black hole solutions in bumblebee gravity and their thermodynamic properties may offer useful insights into potential observational signatures of Lorentz symmetry breaking.
		
		The search for black hole solutions in Lorentz-violating extensions of gravity has attracted considerable attention. In 2017, Casana et al. reported the first exact Schwarzschild-like black hole solution in bumblebee gravity~\cite{Casana:2017jkc}. Since then, a wide variety of solutions have been explored, including traversable wormholes~\cite{Ovgun:2018xys}, Schwarzschild–AdS-like black holes~\cite{Maluf:2020kgf}, slowly rotating Kerr-like black holes~\cite{Ding:2019mal}, and further generalizations~\cite{Santos:2014nxm,Jha:2020pvk,Filho:2022yrk,Xu:2022frb,Ding:2023niy,Liu:2024axg,Bailey:2025oun, Li:2025bzo,Li:2025tcd,Belchior:2025xam,Chen:2025ypx,AraujoFilho:2025rvn,AraujoFilho:2024ykw,Delhom:2019wcm}. Additional black hole solutions have also been found in related SME-inspired frameworks, including models with a nonminimally coupled Kalb–Ramond field possessing a nonzero VEV~\cite{Kalb:1974yc, Altschul:2009ae,Lessa:2019bgi,Kumar:2020hgm,Yang:2023wtu,Duan:2023gng,Do:2020ojg,Liu:2024oas,Liu:2025fxj}. The properties  of these solutions have been extensively investigated~\cite{Guo:2023nkd,Du:2024uhd,Kuang:2022xjp,Oliveira:2018oha,Liu:2022dcn,Zhang:2023wwk,Liu:2024oeq,      Gomes:2018oyd,Kanzi:2019gtu,Gullu:2020qzu,Oliveira:2021abg,Kanzi:2021cbg,Hosseinifar:2024wwe,Liu:2024lve,Deng:2025uvp,Ovgun:2018ran,Sakalli:2023pgn,Mangut:2023oxa,Uniyal:2022xnq,Gu:2025lyz,Lai:2025nyo,Xia:2025yzg}.  
		
		In much of the literature, the bumblebee vector field is assumed to have only a single nonvanishing component. The aim of this work is to derive exact black hole solutions in which the bumblebee vector field has two independent nonzero components, with the VEV being either lightlike or spacelike. We first present new Schwarzschild-like and Schwarzschild–(A)dS-like solutions. Furthermore, by including a nonminimal coupling between the electromagnetic field and the bumblebee vector, we obtain new  charged black hole solutions.   These solutions are characterized by two Lorentz-violating parameters. The bumblebee field strength does not vanish for specific parameter choices. Our solutions demonstrate that Lorentz-violating corrections persist even for a lightlike VEV. In this case,  the Schwarzschild-like and Reissner–Nordström (RN)-like black holes still exhibit asymptotic behavior that deviates from Minkowski spacetime at spatial infinity. Moreover, our thermodynamic analysis reveals a key distinction between spacelike and lightlike VEVs. For spacelike VEVs, our solutions reproduce the known discrepancy—first reported in Ref.~\cite{An:2024fzf} within the Iyer–Wald formalism—between the black hole entropy and the Wald entropy. In contrast, for the lightlike VEV case studied here, we find that the two entropies are identical.
		
		The structure of this paper is organized as follows. Section~\ref{SS2} reviews the framework of bumblebee gravity. In Sec.~\ref{SS3}, we present static, spherically symmetric black hole solutions with lightlike and spacelike VEVs. In Sec.~\ref{SS4}, by introducing a nonminimal coupling between the electromagnetic and bumblebee fields, we derive charged black hole solutions. In Sec.~\ref{SS5}, we analyze the thermodynamic properties of the black holes in Sec.~\ref{SS4} using the Iyer–Wald formalism. Finally, Sec.~\ref{SS6} summarizes our results and provides further discussion.
		%==========================
		
		\section{ EINSTEIN-BUMBLEBEE THEORY }\label{SS2}
		As discussed in the preceding section, the bumblebee model extends general relativity by introducing a vector field, the bumblebee field, which couples nonminimally to gravity. The vector field $B_{\mu}$ acquires a nonzero VEV through a prescribed potential, leading to spontaneous Lorentz-symmetry breaking in the gravitational sector. The action is given by~\cite{Kostelecky:2003fs}

		\begin{align}
			S={}&\int d^{4}x\sqrt{-g}\bigg[
			\frac{1}{2\kappa}\left(R-2\Lambda\right)
			+\frac{\xi}{2\kappa}B^{\mu}B^{\nu}R_{\mu\nu} \notag\\
			&-\frac{1}{4} B_{\mu\nu}B^{\mu\nu}
			-V(B^{\mu}B_{\mu}\pm b^{2}) \bigg]
			+\int d^{4}x\sqrt{-g}\mathcal{L}_{M},
			\label{S1}
		\end{align}
		where $\Lambda$ denotes the cosmological constant, $ \kappa=8\pi G/c^4$ is the gravitational coupling constant, and $\xi$ represents the nonminimal coupling constant between gravity and the bumblebee field. Analogous to the electromagnetic field, the bumblebee field strength is given by
		\begin{eqnarray}
			B_{\mu\nu}&=&\partial_{\mu}B_\nu-\partial_{\nu}B_{\mu}.
			\label{BB}
		\end{eqnarray}
		The potential $V$, which may take the functional form $V(B^{\mu}B_{\mu} \pm b^{2})$, endows the bumblebee field $B_{\mu}$ with a nonvanishing vacuum expectation value. This form imposes the constraint $B^{\mu}B_{\mu} = \mp b^{2}$, resulting in a nonzero vacuum configuration of $B_{\mu}$ that can be spacelike, timelike, or lightlike.  Thus, the field $B_{\mu}$ acquires a vacuum expectation value $\langle B_{\mu}\rangle = b_{\mu}$, where $b_{\mu}$ depends on the spacetime coordinates and satisfies $b^{\mu}b_{\mu}=\mp b^{2}=\mathrm{const}$. For convenience, we define 
	$X = B^{\mu}B_{\mu}\pm b^{2}, 	V' = \frac{\partial V}{\partial X} ,$
		which will be used in the subsequent analysis.

	Next, we briefly discuss the dimensions of the parameters $b$ and $\xi$. 
	In natural units $c=\hbar=1$, the action is dimensionless and the 
	Lagrangian density has dimension $L^{-4}$. 
	From the definition
	$B_{\mu\nu}=\partial_\mu B_\nu-\partial_\nu B_\mu,$
	since $[\partial_\mu]=L^{-1}$, the kinetic term $-\frac14 B_{\mu\nu}B^{\mu\nu}$ implies $[B_\mu]=L^{-1}$. 
	Consequently, the vacuum expectation value $b$ appearing in the potential 
	$V(B^\mu B_\mu\pm b^2)$ has the same dimension as $B_\mu$, namely $[b]=L^{-1}$. 
	For the nonminimal coupling term
	$\frac{\xi}{2\kappa}B^\mu B^\nu R_{\mu\nu}$,
	dimensional consistency then requires $[\xi]=L^{2}$.
	
	The gravitational field equations in the framework of bumblebee gravity are obtained by varying the action~\eqref{S1} with respect to the metric tensor $g^{\mu\nu}$:
	\begin{eqnarray}
			G_{\mu\nu} + \Lambda g_{\mu\nu} &=& \kappa T^{B}_{\mu\nu} \, ,
			\label{modified1}
		\end{eqnarray}
		where the effective energy-momentum tensor for the bumblebee filed is
		\begin{align}
			T^{B}_{\mu\nu}={}&
			\frac{\xi}{\kappa}\bigg[
			\frac{1}{2}B^{\alpha}B^{\beta}R_{\alpha\beta}g_{\mu\nu}
			-B_{\mu}B^{\alpha}R_{\alpha\nu}
			-B_{\nu}B^{\alpha}R_{\alpha\mu} \notag\\
			&+\frac{1}{2}\nabla_{\alpha}\nabla_{\mu}\left(B^{\alpha}B_{\nu}\right)
			+\frac{1}{2}\nabla_{\alpha}\nabla_{\nu}\left(B^{\alpha}B_{\mu}\right) \notag\\
			&-\frac{1}{2}\nabla^{2}\left(B_{\mu}B_{\nu}\right)
			-\frac{1}{2}g_{\mu\nu}\nabla_{\alpha}\nabla_{\beta}
			\left(B^{\alpha}B^{\beta}\right)\bigg] \notag\\
			&+2V'B_{\mu}B_{\nu}
			+B_{\mu}^{\ \alpha}B_{\nu\alpha}
			-\left(V+ \frac{1}{4}B_{\alpha\beta}B^{\alpha\beta}\right)g_{\mu\nu}.
			\label{fa}
		\end{align}
		
		For computational convenience, we can express the field equations in the following form:
		\begin{eqnarray}
			R_{\mu\nu}&=&\Lambda g_{\mu\nu}+ \kappa	\mathcal{T}_{\mu\nu} ,
			\label{modified0}
		\end{eqnarray}
		where
		\begin{align}
			\kappa\mathcal{T}_{\mu\nu}={}&
			\kappa\bigg[
			V'\left(2 B_{\mu}B_{\nu}-b^{2}g_{\mu\nu}\right)
			+B_{\mu}^{\ \alpha}B_{\nu\alpha}
			+V g_{\mu\nu} \notag\\
			&-\frac{1}{4}B_{\alpha\beta}B^{\alpha\beta}g_{\mu\nu}
			\bigg]
			+\xi\bigg[
			\frac{1}{2}B^{\alpha}B^{\beta}R_{\alpha\beta}g_{\mu\nu} \notag\\
			&-B_{\mu}B^{\alpha}R_{\alpha\nu}
			-B_{\nu}B^{\alpha}R_{\alpha\mu}
			+\frac{1}{2}\nabla_{\alpha}\nabla_{\mu}
			\left(B^{\alpha}B_{\nu}\right) \notag\\
			&+\frac{1}{2}\nabla_{\alpha}\nabla_{\nu}
			\left(B^{\alpha}B_{\mu}\right)
			-\frac{1}{2}\nabla^{2}\left(B_{\mu}B_{\nu}\right)
			\bigg] .
			\label{RR1}
		\end{align}
		Similarly, by varying the action \eqref{S1} with respect to the bumblebee vector field, we can obtain 
		\begin{eqnarray}
			\nabla_{\mu}B^{\mu\nu}-2\left( V'B^{\nu}-\frac{\xi}{2\kappa}B_\mu R^{\mu\nu} \right)=0. \label{BB1}
		\end{eqnarray}
\section{ SPHERICALLY SYMMETRIC BLACK HOLE SOLUTIONS }\label{SS3}
We consider the metric ansatz for a static and spherically symmetric spacetime, which is expressed as
\begin{eqnarray}
	{d}{s}^{2}&=&-A(r) {dt}^{2}+S(r
	){dr}^{2}+r^{2} {~d}\Omega ^{2},
	\label{qdc}
\end{eqnarray}
where ${~d} \Omega ^{2}={~d} \theta^{2}+ \sin ^{2} \theta {d} \varphi^{2}$.

Specifically, we consider the effects induced by Lorentz symmetry breaking when the bumblebee field $B_{\mu}$ remains frozen at its VEV $b_{\mu}$. A similar assumption was adopted in Ref.~\cite{Bertolami:2005bh}. In this way, the bumblebee field is fixed as
\begin{eqnarray}B_{\mu}&=&b_{\mu}.\end{eqnarray}
We consider a more general background $b_\mu$ with the form~\cite{Xu:2022frb}
\begin{eqnarray}b_{\mu}&=&\left(b_{t}(r),b_{r}(r),0,0\right).\end{eqnarray}
Utilizing the aforementioned condition $b_{\mu}b^{\mu}=b^{2}=const$, we can derive
\begin{eqnarray}b_r(r)&=&\sqrt{b^2 S(r)+\frac{b_t^2(r)S(r)}{A(r)}}\label{btr}.\end{eqnarray}
When $b \neq 0$, the bumblebee field $B_\mu$ corresponds to a spacelike vector field, 
whereas for $b =  0$, $B_\mu$ is lightlike.

\subsection{Case A: $V(X)=\frac{\lambda}{2}X^2$ and $ \Lambda=0$ } \label{3.1A}
In the absence of a cosmological constant, we impose the vacuum conditions $V=0$ and $V'=0$~\cite{Casana:2017jkc,Kostelecky:1989jw}.  
A simple example of a potential satisfying these conditions is the smooth quadratic form
\begin{eqnarray}
	V(X) &=& \frac{\lambda}{2} X^{2} \, ,
	\label{VV}
\end{eqnarray}
where $\lambda$ is a constant. This form is identical to the Higgs-type potential and is related to the mass structure of the theory~\cite{Kostelecky:1989jw}. In this case, $V$ makes no contribution to the field equations. Other choices, such as $V(X)=\tfrac{\lambda}{2}X^{n}$ ($n \geq 3$), likewise do not contribute, and the corresponding solutions remain consistent with the quadratic case $V(X)=\tfrac{\lambda}{2}X^{2}$.  

Substituting Eqs.~\eqref{qdc}--\eqref{VV} together with the above conditions into the field equations~\eqref{modified0}--\eqref{BB1}, the solution is obtained as
\begin{eqnarray}
	A(r)&=&1-\frac{2 M}{r},\label{M11}\\
	S(r)&=&\frac{1+\ell_1+\ell_2}{A(r)},\label{M12}\\
	b_{t}(r)&=&\alpha,\label{bt0}\\
	b_r(r)&=&\sqrt{b^2 S(r)+\frac{(1+\ell_1+\ell_2)b_t^2(r)}{A(r)^2}}.
\end{eqnarray}
Here $\ell_{1}=\xi b^{2}$ corresponds to the Lorentz-violating parameter frequently considered in previous works~\cite{Casana:2017jkc,Ovgun:2018xys,Liu:2024axg}, while $\ell_{2}=\xi \alpha^{2}$ represents a new Lorentz-violating parameter introduced in this paper.

	Physically, these two parameters represent two independent contributions to Lorentz symmetry breaking associated with different components of the background bumblebee field. 
	The parameter $\ell_1=\xi b^{2}$ is determined by the vacuum expectation value $b$ of the bumblebee field and corresponds to the Lorentz-violating parameter commonly discussed in the literature. 
	On the other hand, $\ell_2=\xi\alpha^{2}$ arises from the nonvanishing temporal component $b_t(r)=\alpha$ of the bumblebee field in our solution. 
	Since these two contributions originate from different components of the vector field, they represent two independent sources of Lorentz violation in the model.

	In the Schwarzschild-like solution derived here, the parameters $\ell_1$ and $\ell_2$ may appear in the combined form $\ell_1+\ell_2$, reflecting that the effective modification of the gravitational field equations depends on a specific combination of the components of the bumblebee field. 
	However, the two parameters are not completely degenerate. 
	As shown in the following sections, when the electric charge and the cosmological constant are included, the metric functions contain $\ell_1$ and $\ell_2$ in different combinations. 
	Moreover, the thermodynamic quantities of the black hole receive distinct corrections from these two parameters. 
	These results indicate that $\ell_1$ and $\ell_2$ encode different physical effects of Lorentz symmetry breaking in spacetime.

In principle, the Lorentz-violating parameter $\ell$ is constrained to be very small by observations. 
For instance, Solar-system tests of the Schwarzschild-like bumblebee black hole metric lead to the bound 
$|\ell|<6.2\times10^{-13}$~\cite{Casana:2017jkc}.

	More recently, the Lorentz-violating parameter has also been constrained in strong-gravity environments. 
	By analyzing the orbital precession of the S2 star around the Galactic-center supermassive black hole Sgr~A*, the parameter $\ell$ in the Schwarzschild-like bumblebee black hole spacetime was constrained to be 
	$\ell = -8.01\times10^{-5} {}^{+2.77\times10^{-4}}_{-2.09\times10^{-4}}$ (uniform prior) and 
	$\ell = 1.00\times10^{-5} {}^{+2.90\times10^{-4}}_{-2.91\times10^{-4}}$ (Gaussian prior) 
	at the $1\sigma$ confidence level~\cite{QiQi:2026zwp}. 
	In addition, gravitational-wave observations also provide potential probes of Lorentz violation, indicating that the parameter combination $\xi {b_t}^2$ is constrained to be $\lesssim 10^{-15}$ within the bumblebee gravity framework~\cite{Lai:2025nyo}.

 There exists a special case when $\xi= \kappa/2$, in which the $t$-component of the bumblebee field in Eq.~\eqref{bt0} takes the form
\begin{eqnarray}
	b_{t}(r) = \alpha + \frac{\beta}{r},
	\label{bt1}
\end{eqnarray}
and consequently the bumblebee field strength becomes nonvanishing,
\begin{eqnarray}
	B_{r t}=-B_{t r} =-\frac{\beta}{r^2}.
\end{eqnarray}
Another important feature of this solution is the physical singularity, which can be investigated via the Kretschmann scalar $K$, constructed from the Riemann tensor:
\begin{eqnarray}
	K &=& R_{\alpha\beta\delta\gamma}R^{\alpha\beta\delta\gamma} \\ \nonumber
	&=&\frac{4 \left((\ell_1+\ell_{2})^2 r^2+4(\ell_1+\ell_2) M r+12 M^2\right)}{(1+\ell_1+\ell_{2})^2 r^6}.
\end{eqnarray}
From this expression we see that the singularity occurs only at $r=0$, which is analogous to the singularity structure of the Schwarzschild black hole.

Compared with the black hole solution for a spacelike bumblebee field with only a nonvanishing $r$-component~\cite{Casana:2017jkc}, the solution obtained here involves two Lorentz-violating parameters. When $\ell_{2}=0$, our black hole solution reduces to that of Ref.~\cite{Casana:2017jkc}. At spatial infinity, the metric takes the form
\begin{eqnarray}
	ds^{2} = -dt^{2} + (1 +\ell_{1}+\ell_{2})\,dr^{2} + r^{2} d\Omega^{2}.
\end{eqnarray}
By performing the coordinate transformation $dr = \sqrt{1/(1 +\ell_{1}+\ell_{2})}\, d\hat{r}$, the asymptotic metric becomes
\begin{eqnarray}
	ds^{2} = -d{t}^{2} + d\hat{r}^{2} + \frac{1}{1+\ell_{1}+\ell_{2}}\,\hat{r}^{2} d\Omega^{2}.\label{tr1}
\end{eqnarray}
This shows that the temporal and radial sectors coincide with those of Minkowski spacetime in spherical coordinates, while the angular part acquires a constant factor $1/(1+\ell_{1}+\ell_{2})$. And the Ricci scalar evaluates to
\begin{eqnarray}
	R = \frac{2(\ell_{1}+\ell_{2})}{(1+\ell_{1}+\ell_{2})r^{2}}.
\end{eqnarray}
Hence, the spacetime is not asymptotically Minkowskian. From a more rigorous perspective, for asymptotically flat spacetimes one usually writes $r=\sqrt{x^{2}+y^{2}+z^{2}}$. As $r\to\infty$ along either timelike or lightlike directions, the metric takes the asymptotic form  
\begin{eqnarray}
	g_{\mu\nu} &=& \eta_{\mu\nu} + \mathcal{O}(r^{-1}),
\end{eqnarray}
so that the deviation from Minkowski spacetime  in the radial component is exactly $(\ell_{1}+\ell_{2})$, i.e., a constant, rather than decaying as $\mathcal{O}(r^{-1})$. 
This shows that the spacetime is not asymptotically flat.

For the case of a lightlike bumblebee field with $b = 0$, we have $\ell_{1} = 0$, and the metric functions take the form
\begin{eqnarray}
	A(r) &=& 1 - \frac{2M}{r}, \\
	S(r) &=& \frac{1 + \ell_{2}}{A(r)} .
\end{eqnarray}
This still represents a Schwarzschild-like black hole solution. It is evident that the spacetime remains not asymptotically flat.

\subsection{Case B: $V(X)=\frac{\lambda}{2}X$ and $\Lambda\neq0$}\label{3.1B}
Next, we extend our analysis to the case with a nonvanishing cosmological constant, aiming to obtain an exact analytical black hole solution within this framework. A simple and convenient choice for the potential is a linear function~\cite{Duan:2023gng}:
\begin{eqnarray}
	V(X) &=& \frac{\lambda}{2}X,
	\label{L2}
\end{eqnarray}
where $\lambda$ is a Lagrange multiplier field~\cite{Bluhm:2007bd}. The equation of motion obtained by varying with respect to $\lambda$ enforces the vacuum condition $X = 0$, which  implies $V = 0$ for any $\lambda$ on shell. This construction effectively freezes fluctuations about the potential minimum, thereby allowing an efficient extraction of the essential physics~\cite{Kostelecky:1989jw}.
However, unlike the vacuum condition in Case A, here $V^{\prime} \neq 0$ whenever $\lambda \neq 0$, so that the potential $V$ continues to contribute to the field equations.

Substituting Eqs.~\eqref{qdc}--\eqref{btr}, together with the above conditions and the chosen form of the potential $V$ given in Eq.~\eqref{L2}, into the field equations~\eqref{modified0}--\eqref{BB1},  we can derive the corresponding field equations.
An analytical solution exists if and only if
\begin{eqnarray}
	\Lambda &=& \frac{\kappa \lambda}{\xi}(1+\ell_1),
	\label{la}
\end{eqnarray}
which follows directly from the field equations.
In this case, a Schwarzschild–(A)dS–like black hole solution featuring two Lorentz-violating parameters is obtained:
\begin{eqnarray}
	A(r)&=&1-\frac{2 M}{r}- \frac{(1+\ell_1+\ell_{2})}{3(1+\ell_1)}\Lambda  r^2 ,\label{M21} \\
	S(r)&=&\frac{1+\ell_1+\ell_2}{A(r)},
	\label{M22}\\
	b_{t}(r)&=&\alpha,\\
	b_r (r)&=&\sqrt{b^2 S(r)+\frac{(1+\ell_1+\ell_2)b_t^2(r)}{A(r)^2}}.
\end{eqnarray}
Compared with the Schwarzschild–(A)dS black hole for a spacelike bumblebee field with only a nonvanishing $r$-component~\cite{Ovgun:2018xys}, the solution obtained here involves two Lorentz-violating parameters. When $\ell_{2}=0$, our solution reduces to that of Ref.~\cite{Ovgun:2018xys}. We can analyze the asymptotic behavior of the metric, taking into account both signs of the effective cosmological constant. After performing a coordinate transformation analogous to Eq.~\eqref{tr1}, the metric can be written in the asymptotic form:
\begin{align}
	ds^{2}={}& -(1-\Lambda_e \hat{r}^2)\,dt^{2}
	+\frac{d\hat{r}^{2}}{1-\Lambda_e \hat{r}^2} \notag\\
	&+\frac{\hat{r}^{2}}{1+\ell_{1}+\ell_{2}}\,d\Omega^{2},
\end{align}
where
\begin{eqnarray}
	\Lambda_e &=& \frac{\Lambda}{(1+\ell_1)}.
\end{eqnarray}
For $\Lambda_e>0$, the spacetime is asymptotically dS and possesses a cosmological horizon located at $\hat{r} \sim 1/\sqrt{\Lambda_e}$. For $\Lambda_e<0$, the spacetime is asymptotically AdS.
The constant factor $1/(1+\ell_{1}+\ell_{2})$ in the angular sector corresponds to a uniform rescaling of the boundary sphere. As a result, the boundary conformal metric is given by
\begin{eqnarray}
	ds_{\rm bdry}^{2} &=& -d\tau^{2} + \frac{1}{1+\ell_{1}+\ell_{2}}\, d\Omega^{2},
\end{eqnarray}
where $\tau$ denotes the time coordinate induced on the conformal boundary. Therefore, up to this constant angular rescaling, the spacetime preserves its asymptotic (A)dS structure.

For the case of a lightlike bumblebee field with $b=0$, we have
\begin{eqnarray}
	A(r)&=&1-\frac{2 M}{r}- \frac{(1+\ell_{2})}{3}\Lambda  r^2 , \\
	S(r)&=&\frac{1+\ell_2}{A(r)},
\end{eqnarray}
which shows that corrections from Lorentz-violating parameters remain.

		Before concluding this subsection, we briefly clarify the role of the
		bumblebee potential in the two classes of solutions considered in this work.
		To obtain exact analytical black hole solutions, different forms of the
		potential are adopted for the cases $\Lambda=0$ and $\Lambda\neq0$.
		In particular, the $\Lambda=0$ solution is derived from the quadratic
		potential $V=\frac{\lambda}{2}X^{2}$, while the $\Lambda\neq0$ solution
		presented above corresponds to the linear potential $V=\frac{\lambda}{2}X$.
		We emphasize that the thermodynamic properties studied in 
		Sec.~5 are derived entirely from this $\Lambda\neq0$ solution obtained
		with the linear potential $V=\frac{\lambda}{2}X$. The limit $\Lambda\to0$ corresponds 
	to $\lambda\to0$. In this limit the derivative of the potential 
	$V'=\lambda/2$ also vanishes, so that the resulting field equations 
	reduce to the same form as those in the $\Lambda=0$ case derived 
	from the quadratic potential. Consequently,
		in the limit $\Lambda\to0$, the black hole solution obtained from the
		linear potential smoothly reduces to the $\Lambda=0$ solution, and the
		corresponding thermodynamic structure consistently approaches that of the
		$\Lambda=0$ case. Therefore, the $\Lambda=0$ solution can be interpreted
		as the zero-pressure limit of the extended phase space thermodynamics.
	
\section{CHARGED SPHERICALLY SYMMETRIC BLACK HOLE SOLUTIONS }\label{SS4}
We consider the matter sector to be described by an electromagnetic field that is nonminimally coupled to the bumblebee vector field~\cite{Liu:2024axg,Lehum:2024ovo}. The corresponding Lagrangian density is given by
\begin{eqnarray}
	\mathcal{L}_{M} &=& -\frac{1}{2\kappa} \left( F^{\mu\nu}F_{\mu\nu} + \gamma B^{\mu}B_{\mu} F^{\alpha\beta}F_{\alpha\beta} \right),
	\label{LF}
\end{eqnarray}
where the electromagnetic field strength tensor is defined as
\begin{eqnarray}
	F_{\mu\nu} &=& \partial_{\mu}A_{\nu} - \partial_{\nu}A_{\mu}, \nonumber\\
	A_{\mu} &=& (\phi(r), 0, 0, 0),
	\label{Fa2}
\end{eqnarray}
and $\gamma$ denotes the coupling coefficient between the electromagnetic field and the bumblebee vector field. 
Since $[F_{\mu\nu}]=L^{-2}$ and $[B_\mu]=L^{-1}$, dimensional consistency implies that 
	$[\gamma]=L^{2}$.
The gravitational field equations in the framework of bumblebee gravity can be derived by varying the action \eqref{S1} with respect to the metric tensor $g^{\mu\nu}$:
\begin{eqnarray}
	G_{\mu\nu}+\Lambda g_{\mu\nu}= \kappa T^{M}_{\mu\nu}+ \kappa T^{B}_{\mu\nu} ,
	\label{modified}
\end{eqnarray}
where
\begin{align}
	T^{M}_{\mu\nu}=\frac{1}{\kappa}\bigg[
	&(1+\gamma b^2)
	\left(2F_{\mu\alpha}F_{\nu}^{\alpha}
	-\frac{1}{2}g_{\mu\nu}F^{\alpha\beta}F_{\alpha\beta}\right) \notag\\
	&+\gamma B_{\mu}B_{\nu}F^{\alpha\beta}F_{\alpha\beta}
	\bigg].
	\label{Tm}
\end{align}
For computational convenience, we can express the field equations as the following form:
\begin{eqnarray}
	R_{\mu\nu}=\Lambda g_{\mu\nu}+ \kappa\mathcal{T}^{M}_{\mu\nu}+\kappa\mathcal{T}^{B}_{\mu\nu} ,
	\label{modified2}
\end{eqnarray}
where
\begin{eqnarray}
	\mathcal{T}^{M}_{\mu\nu}&=&{T}^{M}_{\mu\nu}-\frac{1}{2}{T}^{M}g_{\mu\nu},\nonumber\\
	\mathcal{T}^{B}_{\mu\nu}&=&{T}^{B}_{\mu\nu}-\frac{1}{2}{T}^{B}g_{\mu\nu}.
	\label{RR}
\end{eqnarray}
By varying the action \eqref{S1} with respect to the bumblebee vector field and the electromagnetic field, we can obtain the equations of motion for the corresponding  fields:
\begin{align}
	0={}&\nabla_{\mu}B^{\mu\nu}
	-2\left(
	V'B^{\nu}
	-\frac{\xi}{2\kappa}B_\mu R^{\mu\nu}
	\right. \notag\\
	&\left.
	-\frac{\gamma}{2\kappa} B^{\nu}F^{\alpha\beta}F_{\alpha\beta}
	\right),\label{BB}\\
	0={}&\nabla_{\mu}\left(
	F^{\mu\nu}+\gamma B^{\alpha}B_{\alpha}F^{\mu\nu}
	\right).
	\label{FF}
\end{align}
By substituting the potential $V$ from Eq.~\eqref{VV} together with the ansatz given in Eqs.~\eqref{qdc}–\eqref{btr} and \eqref{Fa2} into the field equations~\eqref{modified2}–\eqref{FF}, and taking the coupling parameter to be $\gamma = \tfrac{\xi}{2+\ell_1}$, we obtain the charged black hole solution as
\begin{eqnarray}
	A(r)&=&1-\frac{2 M}{r}+\frac{ 2(1+\ell_1+\ell_{2})Q_0^2}{(2+\ell_1) r^2},\label{Q11}\\
	S(r)&=&\frac{1+\ell_1+\ell_2}{A(r)},\label{Q12}\\
	b_{t}(r)&=&\alpha,\label{qb}\\
	b_r(r)&=&\sqrt{b^2 S(r)+\frac{(1+\ell_1+\ell_2)b_t^2(r)}{A(r)^2}},\\ 
	\phi(r)&=&-\frac{Q_{0}\sqrt{1+\ell_1+\ell_{2}}}{r}\label{ph}.
\end{eqnarray}
This solution is similar to the RN solution.
Similarly, for the special case of $\xi= \kappa/2$, the $t$-component of the bumblebee field $b_{t}$ in Eq.~\eqref{qb} takes the form given in Eq.~\eqref{bt1}, leading to a nonvanishing bumblebee field strength.
Through the modified Maxwell equations, the conserved current takes the form  
\begin{eqnarray}
	J^{\nu} = \nabla_{\mu}\left(F^{\mu\nu} + \gamma B^{\alpha} B_{\alpha} F^{\mu\nu}\right) \, .
\end{eqnarray}
Consequently, the conserved electric charge $Q$ can be expressed as 
\begin{align}
	Q={}& -\frac{1}{4\pi}\int_{\Sigma} d^{3}x\,
	\sqrt{\gamma^{(3)}}\,n_{\mu} J^{\mu} \notag\\
	={}& -\frac{1}{4\pi}\int_{\partial\Sigma}
	d\theta\,d\phi\,\sqrt{\gamma^{(2)}}\,n_{\mu}\sigma_{\nu}
	\bigg(F^{\mu\nu} \notag\\
	&\hspace{2.4cm}
	+\frac{\xi}{2+\ell_1}\, B^{\alpha} B_{\alpha} F^{\mu\nu}
	\bigg) \notag\\
	={}& \left(1 + b^{2}\frac{\xi}{2+\ell_1}\right) Q_{0} \notag\\
	={}& \frac{2(1+\ell_1)}{2+\ell_1} \, Q_{0} \, . \label{e2}
\end{align}
Here, $\Sigma$ denotes a three-dimensional spacelike hypersurface with induced metric $\gamma_{ij}^{(3)}$, while its boundary $\partial\Sigma$ is a two-sphere at spatial infinity with induced metric $\gamma_{ij}^{(2)} = r^{2} d\Omega^{2}$. The unit normal vectors are given by $n_{\mu}=(1,0,0,0)$ and $\sigma_{\mu}=(0,1,0,0)$, associated with $\Sigma$ and $\partial\Sigma$, respectively.

Similar to the  RN black hole, this solution possesses two horizons, given by  
\begin{eqnarray}
	r_{\pm}=M \pm \sqrt{M^2-\frac{2 (1+\ell_1+\ell_{2})}{2+\ell_1}\,Q_0^2} .
\end{eqnarray}
It follows that the mass and charge parameters of the black hole must satisfy
\begin{eqnarray}
	\frac{Q_0^2}{M^2} \leq \frac{2+\ell_1}{2(1+\ell_1+\ell_{2})} .
\end{eqnarray}
The Kretschmann scalar $K$ is given by:
\begin{align}
	K &= R_{\alpha\beta\delta\gamma}R^{\alpha\beta\delta\gamma} \notag \\[4pt]
	&= \frac{4 \big[(\ell_1+\ell_{2})^2 r^2
	+4(\ell_1+\ell_2) M r+12 M^2 \big]}
	{(1+\ell_1+\ell_{2})^2 r^6} \notag \\[6pt]
	&\quad - \frac{16 Q_0^2 (1+\ell_1)}
	{(1+\ell_1+\ell_2)^2 (2+\ell_1)^2 r^8} \notag\\
	&\qquad \times
	\Big[(2+\ell_{1}) r \big((\ell_1+\ell_2) r+12 M\big)
	-14 (1+\ell_1) Q_0^2\Big].
\end{align}
From this expression, we observe that the singularity appears only at $r = 0$, which is analogous to the singularity structure of the RN black hole. When $\ell_{2} = 0$, our black hole solution reduces to that obtained in Ref.~\cite{Liu:2024axg}. Furthermore, for $\ell_{2} = \ell_{1} = 0$, the solution  recovers the standard RN black hole.

Following the same reason as in the previous section, for the lightlike case with $\ell_{1}=0$, the metric functions take the form:
\begin{eqnarray}
	A(r)&=&1-\frac{2 M}{r}+\frac{(1+\ell_2)Q_0^2}{r^2}  , \\
	S(r)&=&\frac{1+\ell_2}{A(r)}.
\end{eqnarray}
In this case, the coupling constant between the electromagnetic field and the bumblebee field becomes $\gamma=\xi/2$, while the charge satisfies $Q=Q_0$.

Similarly, in the presence of a cosmological constant, an analytical solution can be obtained provided that the condition~\eqref{la} is satisfied.
By substituting the potential~\eqref{L2} together with the condition~\eqref{la} into the field equations~\eqref{modified2}--\eqref{FF}, we cab obtain the corresponding charged (A)dS black hole solution:
\begin{align}
	A(r)={}&1-\frac{2 M}{r}
	+\frac{2(1+\ell_1+\ell_2)Q_0^2}{(2+\ell_1) r^2} \notag\\
	&-\frac{(1+\ell_1+\ell_{2})}{3(1+\ell_1)}
	\Lambda r^2,\label{Q21} \\
	S(r)={}&\frac{1+\ell_1+\ell_2}{A(r)}.
	\label{Q22}
\end{align}
The expressions of $\phi(r)$ and $b_{\mu}$ are consistent with those used in Eqs.~\eqref{qb}–\eqref{ph}. This solution is similar to the RN–(A)dS black hole solution. When $\alpha=0$, our black hole solution reduces to that of Ref.~\cite{Liu:2024axg}. For the case $b=0$, which corresponds to a lightlike bumblebee field, the black hole solution takes the following form:
\begin{eqnarray}
	A(r)&=&1-\frac{2 M}{r}+\frac{ (1+\ell_2)Q_0^2}{ r^2}-\frac{(1+\ell_{2})}{3}\Lambda  r^2 ,\label{Q21} \\
	S(r)&=&\frac{1+\ell_2}{A(r)}.
	\label{Q23}
\end{eqnarray}
As an extension of these black hole solutions, the $n$-dimensional case is discussed in ~\ref{AAA}. 
\section{THERMODYNAMICS}\label{SS5}
As a cornerstone of black hole physics, black hole thermodynamics offers a unique probe of quantum phenomena in curved spacetime~\cite{Hawking:1975vcx,Gibbons:1976ue,Bardeen:1973gs,Hawking:1976de}. In our setup, the bumblebee field is nonminimally coupled to gravity, leading to corrections that go beyond general relativity. This implies that black hole thermodynamics needs to be reconsidered and reformulated. In the following, we employ the Iyer–Wald formalism~\cite{Wald:1993nt,Iyer:1994ys,Iyer:1995kg} to analyze the thermodynamics of the RN–AdS-like black hole solutions obtained in Sec.~\ref{SS4}, with the aim of exploring the effects of the Lorentz-violating parameters on black hole thermodynamics.

Consider the $n$-dimensional spacetime Lagrangian density for gravity, the bumblebee field, and the $U(1)$ vector field, 
\begin{eqnarray}
	\mathbf{L} = L \bm{\epsilon},
\end{eqnarray}
where $\bm{\epsilon}$ denotes the spacetime volume form and $L$ corresponds to the Lagrangian in action \eqref{S1}. Varying the fields $\Phi \equiv \{g_{ab}, B_{a}, A_{a}\}$ gives
\begin{eqnarray}
	\delta\mathbf{L} = \mathbf{E}[\Phi]\delta\Phi + \mathrm{d}\mathbf{\Theta}[\Phi,\delta\Phi],
	\label{eq:LagrangianVariation}
\end{eqnarray}
where $\mathbf{E}[\Phi]$ represents the bulk contribution from the variation (with  $\mathbf{E}[\Phi] =0$ corresponding to the field equations), and $\mathbf{\Theta}[\Phi,\delta\Phi]$ is the presymplectic potential.
It is clear that the action we consider is diffeomorphism invariant; under a field variation $\delta_\xi \Phi = \mathcal{L}_\xi \Phi$, the action remains unchanged. In other words, the Lagrangian density is a $D$-form, and its variation under $\delta_\xi \Phi = \mathcal{L}_\xi \Phi$ is given by
\begin{eqnarray}
	\delta _{\xi}\mathbf{L}=\mathcal{L} _{\xi}\mathbf{L}=\xi \cdot \mathrm{d}\mathbf{L}+\mathrm{d(}\xi \cdot \mathbf{L})=\mathrm{d(}\xi \cdot \mathbf{L}).
\end{eqnarray}
By applying the specific variation given in Eq.~\eqref{eq:LagrangianVariation}, we replace the variation of an arbitrary field with $\delta_\xi \Phi = \mathcal{L}_\xi \Phi$. This substitution yields
\begin{eqnarray}
	\delta_\xi\mathbf{L} = \mathbf{E}[\Phi]\mathcal{L}_\xi\Phi + \mathrm{d}\mathbf{\Theta}[\Phi,\mathcal{L}_\xi\Phi],
\end{eqnarray}
which implies the identity
\begin{eqnarray}
	\mathrm{d}(\xi \cdot \mathbf{L}) = \mathbf{E}[\Phi]\mathcal{L}_\xi\Phi + \mathrm{d}\mathbf{\Theta}[\Phi,\mathcal{L}_\xi\Phi].
\end{eqnarray}
By invoking Noether’s second theorem~\cite{Avery:2015rga,Compere:2019qed}, we conclude that the first term on the right-hand side is an exact form and vanishes on-shell. Consequently, the Noether current $ \textbf{J}_\xi $, which is conserved on-shell (i.e., $\mathrm{d}\mathbf{J}_\xi = 0$), is given by
\begin{eqnarray}
	\mathbf{J}_\xi = \mathbf{\Theta}[\Phi, \mathcal{L}_\xi\Phi] - \xi \cdot \mathbf{L}.
	\label{eq:NoetherCurrent}
\end{eqnarray}
Furthermore, by the algebra Poincaré lemma~\cite{Barnich:2018gdh,Ruzziconi:2019pzd}, this implies that at least locally there exists a $(n-2)$-form, called the Noether charge $\mathbf{Q}_\xi$, such that
\begin{eqnarray}
	\mathbf{J}_\xi = \mathrm{d}\mathbf{Q}_\xi.
	\label{eq_JdQ}
\end{eqnarray}
Building on this expression, taking the variation of the fields in Eq.~\eqref{eq:NoetherCurrent} yields
\begin{align}
	\delta \mathrm{d}\mathbf{Q}_{\xi}&=\delta [\mathbf{\Theta }(\Phi ,\mathcal{L} _{\xi}\Phi )]-\xi \cdot \delta \mathbf{L} \nonumber
	\\
	&=\delta [\mathbf{\Theta }(\Phi ,\mathcal{L} _{\xi}\Phi )]-\mathcal{L} _{\xi}[\mathbf{\Theta }[\Phi ,\delta \Phi ]]+d(\xi \cdot \mathbf{\Theta }[\Phi ,\delta \Phi ]).
\end{align}
This implies that
\begin{align}
	\mathrm{d}\left(
	\delta \mathbf{Q}_{\xi}
	-\xi \cdot \mathbf{\Theta }[\Phi ,\delta \Phi]
	\right)
	={}& \delta [\mathbf{\Theta }(\Phi ,\mathcal{L} _{\xi}\Phi )] \notag\\
	&-\mathcal{L} _{\xi}
	[\mathbf{\Theta }[\Phi ,\delta \Phi ]].
	\label{eq_dkxi}
\end{align}
The right-hand side of this equality is precisely the symplectic current $\bm{\omega}[\delta\Phi,\mathcal{L}_\xi\Phi]$, which corresponds to the surface charge $H_{\xi}$ associated with the field variation $\delta_\xi \Phi = \mathcal{L}_\xi \Phi$. Specifically, it can be expressed as
\begin{eqnarray}
	\delta H_{\xi} =\int_{S_{\infty}}\left({\delta \mathbf{Q}_{\xi}-\xi \cdot \mathbf{\Theta }[\Phi ,\delta \Phi ]}\right)
	\label{eq_deltaH}.
\end{eqnarray}
The term \(\xi \cdot \mathbf{\Theta}[\Phi, \delta \Phi]\) is a boundary term~\cite{Wald:1993nt,Iyer:1994ys}, as is standard in covariant phase space analysis.
From the variation of the action~\eqref{S1}, the presymplectic potential is given by
\begin{align}
	\mathbf{\Theta}[\Phi,\delta\Phi]_{bcd}
	={}& \Big(
	2{E_R}^{a e f h}\nabla_{h}\delta g_{e f}
	- 2\left(\nabla_{h}{E_R}^{a e f h}\right)\delta g_{e f}
	\notag\\
	&- B^{a e}\delta B_{e}
	-\frac{2}{\kappa}
	\left(1+{\gamma}B^{f}B_{f}\right)F^{a e}\delta A_{e}
	\Big) \notag\\
	&\times \varepsilon_{abcd},
\end{align}
where
\begin{align}
	{E_R}^{abcd} &= \frac{1}{2\kappa}\left(
	X^{abcd} + \xi B^{e}B^{f}{Y_{ef}}^{abcd}
	\right),\\
	X^{abcd} &= g^{a[c}g^{d]b},\\
	{Y_{ef}}^{abcd} &= \frac{1}{2}\left(
	{g_{(e}}^a{g_{f)}}^{[c}g^{d]b}
	- {g_{(e}}^b{g_{f)}}^{[c}g^{d]a}
	\right).
\end{align}
Accordingly, using Eqs. \eqref{eq:NoetherCurrent} and \eqref{eq_JdQ}, we obtain the Noether charge 
\begin{align}
	(\mathbf{Q}_\xi)_{cd}={}& \Big(
	- {E_R}^{ab e f}\nabla_{e}\xi_{f}
	- 2\xi_{e}\nabla_{f}{E_R}^{ab e f} \notag\\
	&-\frac{1}{2}B^{ab}B_{f}\xi^{f}
	-\frac{1}{\kappa}(1+{\gamma}B^{f}B_{f})
	F^{a b}A_{e}\xi^{e}
	\Big)\varepsilon_{abcd},
	\label{QQ}
\end{align}
For a stationary black hole with a bifurcate Killing horizon $S_h$ generated by $\xi_H$, we integrate this Eq.~\eqref{QQ} over a hypersurface $V_r $ that extends from the $S_h$ to another codimension-2 surface $ S_r $. Applying Stokes' theorem, the volume integral (for hypersurface) reduces to a surface integral, yielding
\begin{eqnarray}
	\int_{S_r} (\delta\textbf{Q}_{\xi_H} - \xi_H \cdot \mathbf{\Theta}[\Phi, \delta \Phi]) - \int_{S_h} \delta \textbf{Q}_{\xi_H} = 0.
	\label{eq_FirstLawGen}
\end{eqnarray}
Since the Killing vector $\xi_H$ vanishes on bifurcate surface $S_h$, its contribution to the second term in Eq.~\eqref{eq_FirstLawGen} simplifies. Notably, the properties of $\xi_H$ and $S_h$ allow us to express
\begin{eqnarray}
	\int_{S_h}\delta\mathbf{Q}_{\xi_H} = T_H\delta S+\Phi_\text{bh}\delta Q.
\end{eqnarray}
In the above discussion, the conventional variation $\delta$ acts only on the fields while keeping the cosmological constant fixed. However, in many cases, the cosmological constant is treated as a variable pressure in the thermodynamics of AdS black holes, leading to an extended formulation of black hole thermodynamics~\cite{Kubiznak:2012wp,Dolan:2010ha,Dolan:2011xt,Kubiznak:2016qmn,Wei:2015iwa,Cai:2013qga,Kastor:2009wy,Xiao:2025gle}. To derive the extended first law, a new variation $\tilde{\delta}$ is introduced, which acts on the fields as~\cite{Xiao:2023lap}
\begin{eqnarray}
	\tilde{\delta}\Phi =\delta \Phi +\partial _{\Lambda}\Phi \tilde{\delta}\Lambda,
\end{eqnarray}
incorporating variations of the cosmological constant. Accordingly, the extended version of Eq.~\eqref{eq_FirstLawGen} can be obtained as~\cite{Xiao:2023lap},
\begin{align}
	&\int_{\infty}\left[
	\tilde{\delta}\mathbf{Q}_{\xi_H}
	-\xi_H\cdot\mathbf{\Theta}[\Phi,\tilde{\delta}\Phi]
	\right]
	-\int_{{S_h}}\tilde{\delta}\mathbf{Q}_{\xi_H} \notag\\
	&\qquad =
	\frac{\tilde{\delta}\Lambda}{8\pi}
	\int_{V_r}\xi_H\cdot\epsilon.
	\label{eq_ExFirstLawGen}
\end{align}

	In the covariant phase space formalism, the variation appearing in Eq.~\eqref{eq_deltaH} is defined in the space of field configurations. Our setup, $\alpha$, is treated as a fixed parameter, analogous to the vacuum expectation value of the bumblebee field, $b$, and is therefore not included in the variation of the solution space.
	For the black hole solution considered here, the fields depend on a set of parameters such as $M$ and $Q_0$. 
The field variation can thus be expressed in terms of the variations of these parameters as:
\begin{eqnarray}
	 \delta\Phi = \frac{\partial \Phi}{\partial M}\delta M + \frac{\partial \Phi}{\partial Q_0}\delta Q_0,
	 \end{eqnarray}

	Substituting the explicit solution into Eq.~\eqref{eq_deltaH}, the surface integral determining the conserved charge is evaluated on the sphere at spatial infinity. 
	Due to the asymptotic fall-off behavior of the fields, the contributions proportional to $\delta Q_0$  vanish at spatial infinity, so that the conserved charge only receives a nonvanishing contribution from the $\delta M$ term. After restricting the variation to the solution space, the charge variation becomes integrable. Thus, using Eq.~\eqref{eq_deltaH}, the energy associated with the Killing vector $\partial_t$ is given by
\begin{eqnarray}
	\delta E_W =\frac{ (1+\ell_1) \delta M}{\sqrt{1+\ell_1+\ell_2}} 
	\quad \implies \quad 
	E_W = \frac{ (1+\ell_1) M}{\sqrt{1+\ell_1+\ell_2}},
\end{eqnarray}
where the integration constant has been set to zero. The temperature can be obtained from the surface gravity,
\begin{align}
	T_H = \frac{\kappa}{2\pi}
	= \frac{1}{4\pi r_h\sqrt{1+\ell_1+\ell_2}}
	&\left(
	1 - \frac{ 2(1+\ell_1+\ell_2)Q_0^2}{(2+\ell_1) r_h^2} \right. \nonumber \\
	&\left. - \frac{(1+\ell_1+\ell_{2})}{(1+\ell_1)}\Lambda r_h^2
	\right),
	\label{eq_HawkingTem}
\end{align}
where $r_h$ denotes the  event horizon radius of the black hole. The horizon electric potential is defined by 
$\Phi_{\text{bh}} = -({\xi_H}^{a} A_{a})\big|_{S_h}$, yielding
\begin{eqnarray}
	\Phi_\text{bh}&=&-\frac{Q_{0}\sqrt{1+\ell_1+\ell_{2}}}{r_h}.
\end{eqnarray}
The electric charge is given by
\begin{eqnarray}
	Q=\frac{2(1+\ell_1)}{2+\ell_1} \, Q_{0},
\end{eqnarray}
as displayed in Eq.~\eqref{e2}.
On the other hand, the Wald entropy is given by
\begin{eqnarray}
	S_W =-2\pi \int_{S_h}{{E_R}^{abcd}\epsilon _{ab}\epsilon _{cd}}= \left(1+\frac{\ell_1}{2}\right)\pi r_h^2,
\end{eqnarray}
where $\epsilon _{ab}$ denotes the bi-normal vector associated with the Killing horizon. For the special case $\xi= \kappa/2$ and $b_{t}(r) = \alpha + \beta/{r}$, the thermodynamic results remain consistent.
Within the extended Iyer-Wald formalism, the general form of the first law is given by Eq.~\eqref{eq_ExFirstLawGen}.

However, for the specific gravitational model and black hole solutions considered in this work, the Wald entropy $S_W$ must be replaced by the thermodynamic entropy  
\begin{eqnarray}
	S = (1+\ell_{1})\pi r_{h}^{2},
	\label{eq_ThermS}
\end{eqnarray}
in order for the first law to be satisfied. Specifically, the appropriate form of the first law  is
\begin{align}
	\tilde{\delta}E_W 
	&+ \frac{r^{3}}{6}\sqrt{(1+\ell_{1}+\ell_{2})}\,\tilde{\delta}\Lambda
	- \Phi_{\rm bh}\,\tilde{\delta }Q
	- T_{H}\,\tilde{\delta}S \nonumber\\
	&= \frac{r^{3}}{6}\sqrt{(1+\ell_{1}+\ell_{2})}\,\tilde{\delta}\Lambda 
	- \frac{\tilde{\delta}\Lambda}{8\pi}\,
	\frac{4}{3}\pi\sqrt{(1+\ell_{1}+\ell_{2})} r_{h}^{3}.
	\label{eq:ExtendedFirstLaw}
\end{align}

Clearly, $S_{W}$ and the thermodynamic entropy \eqref{eq_ThermS} do not coincide whenever $\ell_{1} \neq 0$, indicating a mismatch between the Wald entropy and the entropy inferred from the first law. 
A similar discrepancy was reported in Ref.~\cite{An:2024fzf}, which corresponds to the case in our work with $\ell_{1}\neq 0$, $\ell_{2}=0$, $Q_{0}=0$, and $\Lambda=0$. 

Similar entropy mismatches have also been observed in modified gravity theories with nonminimal couplings, such as Horndeski gravity~\cite{Feng:2015oea,Feng:2015wvb}. 
In those theories the scalar field is nonminimally coupled to curvature, while in the bumblebee model the vector field is nonminimally coupled to gravity. 
Such couplings can modify the horizon surface terms appearing in the Iyer--Wald construction
\begin{eqnarray}
	\int_{S_h} \big(\delta \mathbf{Q}_{\xi_H} - \xi_H \cdot \mathbf{\Theta}[\Phi, \delta \Phi]\big), 
	\quad \text{where} \; \xi_H =\partial_t ,
\end{eqnarray}
and lead to deviations between the Wald entropy and the thermodynamic entropy.

	In the present solution the bumblebee field satisfies 
	$B_\mu B^\mu=b^2\neq0$, corresponding to a spacelike vacuum 
	expectation value. The entropy mismatch observed in our 
	results is related to the condition $B_\mu B^\mu=b^2\neq0$ 
	together with the nonminimal coupling between the vector 
	field and the curvature. In this situation, the nonminimal 
	coupling contributes an additional term to the Iyer--Wald 
	surface charge, which leads to a deviation between the 
	thermodynamic entropy and the Wald entropy. 
	The spacelike VEV characterizes the Lorentz-symmetry-breaking 
	background but does not by itself spoil the consistency of 
	the theory. Previous studies of the bumblebee model suggest 
	that stable configurations can exist when the potential 
	enforces the vacuum condition $B_\mu B^\mu=b^2$ and the 
	Hamiltonian remains bounded from below 
	~\cite{Bluhm:2004ep,Bluhm:2008yt}. A detailed perturbative 
	stability analysis of the present solution is beyond the 
	scope of this work and is left for future investigation.

In contrast, in the case of $b = 0$, which corresponds to a lightlike bumblebee field, we have $\ell_{1} = 0$. Consequently, the Wald entropy $S_{W}$ coincides with the thermodynamic entropy $S$,
\begin{eqnarray}
	S_{W} = S = \pi r_{h}^{2}.
\end{eqnarray}

	In this lightlike configuration the invariant satisfies 
	$B_\mu B^\mu=0$, and the additional contribution arising 
	from the nonminimal coupling vanishes. Therefore the 
	Iyer--Wald formalism correctly reproduces the thermodynamic 
	entropy in this case.

The formulation of extended thermodynamics yields
\begin{eqnarray}
	\tilde{\delta}E_W = T_H\tilde{\delta}S_W + V\tilde{\delta} P+\Phi_{\rm bh}\,\tilde{\delta}  Q,
\end{eqnarray}
where the thermodynamic quantities are given by
\begin{align}
	E_W &= \frac{M}{\sqrt{1+\ell_2}}, \\[4pt]
	T_H &= \frac{\kappa}{2\pi}
	= \frac{1}{4\pi r_h \sqrt{1+\ell_2}}
	\left(1 - \frac{2(1+\ell_2) Q_0^2}{r_h^2} \right. \nonumber \\[4pt]
	&\qquad - \left. (1+\ell_2)\Lambda r_h^2 \right),
	\label{eq_HawkingTem2} \\[4pt]
	Q &= Q_0, \\[4pt]
	\Phi_{\text{bh}} &= -\frac{Q_{0}\sqrt{1+\ell_2}}{r_h}, \\[4pt]
	V &= \frac{4}{3}\pi\sqrt{1+\ell_{2}}\, r_{h}^{3}, \\[4pt]
	P &= -\frac{\Lambda}{8\pi}.
\end{align}
Using the scaling relations among the thermodynamic quantities, the Smarr relation is obtained as
\begin{eqnarray}
	\quad E_W=2T_H S_W-2PV+\Phi_\text{bh} Q.
	\label{SM1}
\end{eqnarray}
It follows that, although some thermodynamic quantities of the black hole are modified by the Lorentz-violating parameters, the Smarr relation remains valid.
\section{ CONCLUSIONS }\label{SS6}
In this paper, we investigated black hole solutions in bumblebee gravity, configurations where the bumblebee vector field acquires either lightlike or spacelike VEVs while possessing two independent nonvanishing components.
We first presented Schwarzschild-like and Schwarzschild–(A)dS-like solutions. Furthermore, by introducing a nonminimal coupling between the electromagnetic field and the bumblebee field, we obtained new classes of charged black hole solutions. In comparison with earlier bumblebee black hole solutions possessing only a nonvanishing radial component of the field, the present solutions involve two Lorentz-violating parameters and display non-Minkowskian behavior when the cosmological constant vanishes, but remain asymptotically (A)dS when the cosmological constant is present. In contrast to many previous black hole solutions where the bumblebee field strength vanishes, our solutions admit nonvanishing field strength for specific parameter choices. More importantly, the newly obtained black hole solutions with a lightlike VEV are formally similar to those with a spacelike VEV, and they are still being affected by the Lorentz-violating parameter.

We then analyzed the thermodynamics of our obtained RN–AdS-like black holes using the Iyer–Wald formalism and obtained some interesting results.  For the spacelike VEV case, as in previously studied black holes with a single Lorentz-violating parameter, the thermodynamic entropy $S$ does not coincide with the Wald entropy $S_W$. This discrepancy likely originates from the divergent behavior of the bumblebee field near the horizon. In contrast, for the lightlike VEV case, the thermodynamic entropy $S$ agrees precisely with the Wald entropy $S_W$, confirming that the Iyer–Wald formalism provides a consistent thermodynamic description. Although the Lorentz-violating parameters modify certain thermodynamic quantities relative to the RN–AdS black hole, the Smarr relation remains valid.
\appendix
\section{ A BRIEF DISCUSSION OF BLACK HOLE SOLUTIONS IN $n$ DIMENSIONS  }\label{AAA}
We now consider a generalized action for bumblebee gravity in $n$-dimensional spacetime:
\begin{align}
	S={}& \int d^{n}x \sqrt{-g} \bigg[
	\frac{1}{2\kappa}\left(R - 2\Lambda\right)
	+\frac{\xi}{2\kappa} B^{\mu} B^{\nu} R_{\mu\nu}
	\notag\\
	&-\frac{1}{4} B_{\mu\nu} B^{\mu\nu}
	-V(B^{\mu} B_{\mu} \pm b^{2})
	\bigg]
	+\int d^{n}x \sqrt{-g}\, \mathcal{L}_{M},
	\label{Sn}
\end{align}
the matter Lagrangian density is given by~\eqref{LF}.  

For the spacetime geometry, we adopt the following static, spherically symmetric ansatz:
\begin{eqnarray}
	ds^2 = - A(r)\, dt^2 + S(r)\, dr^2 + r^2 d\Omega_{n-2}^2 ,
	\label{mm}
\end{eqnarray}
where $d\Omega_{n-2}^2$ is the line element of a unit $(n-2)$-sphere.  

The bumblebee field and the $U(1)$ gauge field compatible with static spherical symmetry take the form
\begin{eqnarray}
	b_\mu dx^\mu &=& b_{t}(r)\, dt + b_{r}(r)\, dr, \\
	A_\mu dx^\mu &=& \phi(r)\, dt. \label{Fa}
\end{eqnarray}

First, we consider the case without a cosmological constant, for which we adopt the effective potential chosen in Sec~\ref{3.1A}, we consider the Lagrangian density given in Eq.~\eqref{LF}. Considering the following relation between the coupling constant and other parameters
\begin{eqnarray}
	\gamma = \frac{(n-3)^{2}\,\xi}{\,n-2+\ell_{1}\,},
\end{eqnarray}
we obtain a $n$-dimensional black hole solution:
\begin{eqnarray}
	A(r)&=&1-\frac{\mu}{r^{n-3}}+\frac{ 2(n-3)(1+\ell_1+\ell_{2})Q_0^2}{(n-2+\ell_1) r^{2(n-3)}},\label{Qn1}\\
	S(r)&=&\frac{1+\ell_1+\ell_2}{A(r)},\label{Qn2}\\
	b_{t}(r)&=&\alpha, \label{tn}\\
	b_r(r)&=&\sqrt{b^2 S(r)+\frac{(1+\ell_1+\ell_2)b_t^2(r)}{A(r)^2}},\\
	\phi(r)&=&\frac{Q_{0}\sqrt{1+\ell_1+\ell_{2}}}{r^{n-3}}\label{phn}.
\end{eqnarray}

Similarly, for the special case of $\xi= \kappa/2$,
the $t$-component of the bumblebee field in Eq.~\eqref{tn} takes the form
\begin{eqnarray}
	b_{t}(r) = \alpha + \frac{\beta}{r^{n-3}},
	\label{bt_n}
\end{eqnarray}
and consequently the bumblebee field strength becomes nonvanishing,
\begin{eqnarray}
	B_{rt} = - B_{tr} =- \frac{(n-3)\beta}{r^{n-2}} .
	\label{Brt_n}
\end{eqnarray}
This $1/r^{\,n-2}$ falloff is analogous to the behavior of a Coulomb-type field in $n$-dimensional spacetime. In the case $Q_{0}=0$, the solution naturally reduces to the vacuum case with $\mathcal{L}_{M}=0$.  

Next, we consider the case with a nonvanishing cosmological constant and 
adopt the effective potential chosen in Sec.~\ref{3.1B}.  In order to obtain an exact solution, we introduce the following condition: 
\begin{eqnarray}
	\Lambda = \frac{2\kappa \lambda}{(n-2)\xi}(1+\ell_1).
	\label{lan}
\end{eqnarray}
Under this condition, the corresponding $n$-dimensional black hole solution takes the form
\begin{eqnarray}
	A(r)&=&1-\frac{\mu}{r^{n-3}}+\frac{ 2(n-3)(1+\ell_1+\ell_{2})Q_0^{2}}{(n-2+\ell_1) r^2}\nonumber \\ 
	&-&   \frac{2(1+\ell_1+\ell_{2})}{(n-1)(n-2)(1+\ell_1)}\Lambda  r^2,\label{Qn3}       \\ 
	S(r)&=&\frac{1+\ell_1+\ell_2}{A(r)}.
\end{eqnarray}
The explicit forms of $b_{t}(r)$, $b_{r}(r)$, and $\phi(r)$ are identical to those presented in Eqs.~\eqref{tn}–\eqref{phn}.

		%=========================================================
		\section*{Acknowledgments}
		This work was supported by the National Natural Science Foundation of China (Grants No. 12475056, No. 12475055, No. 12247101 ), 
		the 111 Project (Grant No. B20063), 
		and Gansu Province's Top Leading Talent Support Plan. \\
		~\\
		\textbf{Conflict of Interest}  The authors declare that they have no conflict of interest.
		%=========================================================
		
		%=================================================

		%=============================

		%\Authorfootnote{}
		
		% \newpage
		%\bibliographystyle{scichina}
		%\bibliographystyle{scpma}
		%\bibliography{gwc}
	
\end{document}